# Separable Temporal Convolution plus Temporally Pooled Attention for Lightweight High-performance Keyword Spotting

Shenghua Hu[*], Jing Wang[*], Yujun Wang[†], Wenjing Yang[*]
[*]Beijing Institute of Technology, Beijing, China
E-mail: {hushenghua, wangjing,ywj}@bit.edu.cn  Tel: +86-10-13810150086
[†]Xiaomi Inc., Beijing, China
E-mail: wangyujun@xiaomi.com  Tel: +86-10-18310875062

*Abstract*— Keyword spotting (KWS) on mobile devices generally requires a small memory footprint. However, most current models still maintain a large number of parameters in order to ensure good performance. In this paper, we propose a temporally pooled attention module which can capture global features better than the AveragePool. Besides, we design a separable temporal convolution network which leverages depthwise separable and temporal convolution to reduce the number of parameter and calculations. Finally, taking advantage of separable temporal convolution and temporally pooled attention, a efficient neural network (ST-AttNet) is designed for KWS system. We evaluate the models on the publicly available Google speech commands data sets V1. The number of parameters of proposed model (48K) is 1/6 of state-of-the-art TC-ResNet14-1.5 model (305K). The proposed model achieves a 96.6% accuracy, which is comparable to the TC-ResNet14-1.5 model (96.6%).

## I. Introduction

KWS aims at detecting predefined keywords in audio signals and is widely implemented in hands-free control of mobile devices. Recently, many human-machine interfaces (HCI) rely on KWS, such as Apple Siri [1], Microsoft Cortana, Amazon Alexa [2,3] and Google Assistant [4], etc. These systems exploit powerful neural network models, which usually run in the cloud. Lightweight KWS models enable the development of many novel engineering applications, such as high-real time voice control equipment and operation without Internet coverage. However, it's still a very challenging task to implement a slight and accurate KWS model on mobile devices with limited hardware resources.

Nowadays, there are two mainstreams to realize KWS, one based on large vocabulary continuous speech recognition (LVCSR) [5] and another based on hidden Markov model (HMMS) [6]. The main weaknesses in these methods are huge memory cost and calculations, which makes it difficult to be applied on mobile devices.

With the success of deep learning in various fields, KWS based on neural networks has become popular at present [7,8,9,10]. Most of deep neural networks based KWS considers keyword spotting as an audio classification task, in which each keyword is denoted as a class. All other words is represent as a class named "filler". A deep neural network processes acoustic features and outputs the posteriors of the keywords. The keyword is detected, when the confident score exceeds a threshold. Compared to traditional approach, deep neural networks based KWS shows the small footprint, low computational cost, and high performance. However, previous work still use several hundred thousand parameters to achieve the state-of-the-art performance. We believe that the number of parameters can still be further reduced while the performance maintain not be hurt.

The ResNet based KWS system [7] proposed by Tang and Lin, which applied residual connections [11] to improve network depth, have achieved good performance on the Google Speech Commands data sets [12]. However, because of the number of hidden layers and filters, their best model still has more than 200K parameters and 800M multipliers. To address the problem, a temporal convolutional neural network for real-time KWS [10], which leverages 1D convolution along the temporal dimension to enhance the accuracy and reduce the parameters of network, was proposed. Despite their success, the number of filters make the model not enough subtle due to too many channels.

In this paper we propose a separable temporal convolution neural network (ST-AttNet) with attention for KWS, a model composed by CNN (separable temporal convolution) and attention [13]. The CNN captures local features of a sequence, and the attention module captures global features. Compared to TC-ResNet, The proposed model decrease parameters and multipliers due to separable convolution [14]. Moreover, we demonstrate that adding the Attention mechanism into the CNN structure through a specific way, denoted as temporally pooled attention, can improve the performance of the model. Our contributions are as follows:

- We proposed an efficient neural network structure based on separable temporal convolutions and attention mechanism for lightweight high-performance KWS.
- We proposed a specific way to add attention to the CNN structure, which is able to improve the performance of the model.
- Our ST-AttNet evaluated on Googles Speech Commands dataset achieves the state-of-the-art accuracy of 96.6% with only 48K parameters.

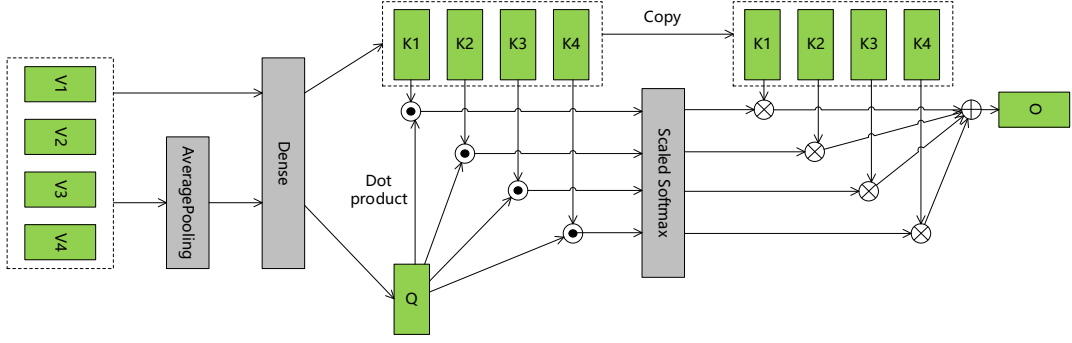

Fig. 1 The figure shows the proposed temporally pooled attention. $v_i$ stands for the *i-th* vector of the temporally pooled attention input.

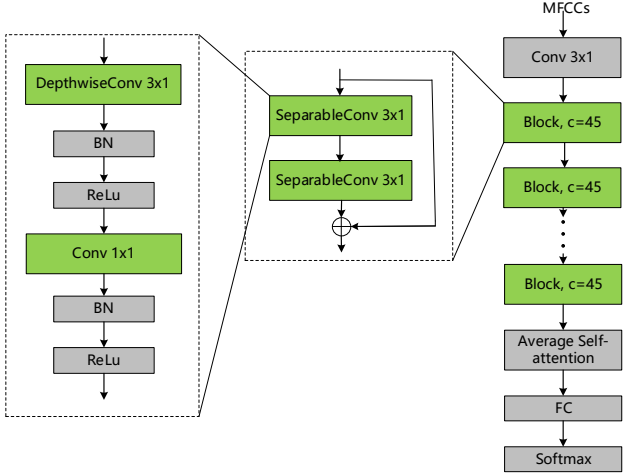

Fig. 2 A schematic of the proposed architecture, with magnified residual block and SeparableConv module. FC denote fully connected layer. 'c' indicates channel size.

## II. PROPOSED METHOD

### A. Data Process

At the beginning of data preprocessing, we apply a band-pass filter of 20Hz/7.8kHz to reduce noise. After that, 40 dimensional Mel-Frequency Cepstrum Coefficient (MFCC) frames are constructed with a 30ms window size and 10ms frame shift. Finally, we feed the MFCC features as input data into the neural network. Correspondingly, we denote the MFCC representation as $I \in \mathbb{R}^{T \times F}$, where F represents the dimension of the MFCC feature, and T denote the number of frames.

### A. Temporally pooled attention

Inspired by success of self-attention mechanism [13] in many tasks, we propose a temporally pooled attention module to improve the performance. A schematic of temporally pooled attention is shown in the Fig. 1.

The original self-attention mechanism can be formulated as follows:

$$\text{Attend}(Q, K, V) = \text{softmax}(\frac{QK^T}{\sqrt{D_k}})V, \quad (1)$$
$$\text{where } Q = UW_q, K = UW_k, V = UW_v.$$

$Q \in \mathbb{R}^{T_u \times D_k}$, $K \in \mathbb{R}^{T_u \times D_k}$, $V \in \mathbb{R}^{T_u \times D_v}$ denote a query matrix, a key matrix, and a value matrix, respectively, and $U \in \mathbb{R}^{T_u \times D_u}$ is the input matrix of the attention module. $T_u$ is the length of the matrix U. $D_k$ is the dimensionality of queries and keys. $D_v$ is the dimensionality of values. Weight matrices $W_q$, $W_k$ and $W_v$ are leveraged to project U into different space.

The above self-attention mechanism input a 2D matrix and output a 2D matrix. However, we demand to reduce 2D input matrix to 1D output vector before Softmax layer. In the previous work [7,8,10], researchers apply an AveragePool layer to achieve the target directly. The advantage of this method is that it has no parameters, but the direct average is equivalent to assigning the same weight to all the features over time series, which is obviously not reasonable. Therefore, we propose a temporally pooled attention module:

$$\text{Attend}(Q, K, V) = \text{softmax}(\frac{QK^T}{\sqrt{D_k}})V, \quad (2)$$
$$\text{where } Q = \text{AveragePool}(U)W, K = UW, V = UW.$$

$W \in \mathbb{R}^{D_u \times D}$ is a shared weight matrix to project input into query, keys and values, inspired Shared Weight Self-Attention [15]. An AveragePool layer is leveraged to transform 2D matrix input into 1D vector i.e. $Q \in \mathbb{R}^D$. So output of temporally pooled attention is a 1D vector.

The module can be extended to a multi-head version [13], which is denoted as follows:

$$\text{Multihead}(Q, K, V) = \text{Cat}(h_1, ..., h_n),$$
$$h_i = \text{Attend}(Q_i, K_i, V_i), \quad (3)$$
$$Q_i = \text{AveragePool}(U)W_i, K_i = UW_i, V_i = UW_i.$$

where n is the number of the heads, $W_i \in \mathbb{R}^{D_u \times (D/n)}$ is the weight matrix of each head, D/n is an integer. So the dimensionalities of the input and the output are the same. After attention, the heads are concatenated together.

Table 1 Parameters used for ST-AttNet4, along with the number of parameters and multipliers. k is the kernel size along the temporal dimension. c is the dimension of the output. d represents the dilation rate.

| Layer | k | c | d | Para. | Mult. |
|---|---|---|---|---|---|
| conv | 3 | 45 | - | 1.9K | 188K |
| res×4 | 3 | 45 | $2^{\lfloor \frac{i}{3} \rfloor}$ | 17.2K | 1.69M |
| avg-att | - | 45 | - | 4.3K | 207K |
| softmax | - | 12 | - | 0.5K | 540 |
| Total | - | - | - | 24K | 2.0M |

Table 2 Parameters used for ST-AttNet4-wide.

| Layer | k | c | d | Para. | Mult. |
|---|---|---|---|---|---|
| conv | 3 | 65 | - | 2.7K | 266K |
| res×4 | 3 | 65 | $2^{\lfloor \frac{i}{3} \rfloor}$ | 35.3K | 3.46M |
| avg-att | - | 65 | - | 8.5K | 428K |
| softmax | - | 12 | - | 0.8K | 780 |
| Total | - | - | - | 48K | 4.1M |

Table 3 Parameters used for ST-AttNet7.

| Layer | k | c | d | Para. | Mult. |
|---|---|---|---|---|---|
| conv | 3 | 45 | - | 1.9k | 188K |
| res×4 | 3 | 45 | $2^{\lfloor \frac{i}{3} \rfloor}$ | 17.2K | 1.69M |
| res×3 | 3 | 45 | - | 12.9K | 1.27M |
| avg-att | - | 45 | - | 4.3K | 207K |
| softmax | - | 12 | - | 0.5K | 540 |
| Total | - | - | - | 37K | 3.3M |

### B. Model Architecture

Following the implementation of temporal convolution [10], we transform the input MFCC into time series feature map at first i.e. the feature dimension is equal to the channels of the input feature map. The input feature map can be denoted as $W \in \mathbb{R}^{D_u \times D}$. As a result, convolutional operations in the model are along the temporal dimension, avoiding stacking many layers to form higher-level features.

Inspired by MobileNet [14], all convolution in the model adopt separable convolution composed by a depthwise convolution applying zero padding with 3×1 kernels and a pointwise convolution with 1×1 kernels. After each depthwise and pointwise convolution layer, there are a batch normalization layer and a ReLU activation unit. None of the convolution layers and fully connected layers have biases, and each batch normalization layer [16] has trainable parameters for scaling and shifting. To expand receptive field over temporal dimension more effciently, dilated convolutions is adopted to all depthwise convolution layer. An exponential sizing schedule [17] is used: at layer i, the dilation is $d = 2^{\lfloor \frac{i}{3} \rfloor}$. The base residual block of proposed model is consisted of 2 separable convolution layers with a residual connect from block input to block output. Unlike standard Temporal convolution, we don't use convolution with stride of two, instead of convolution with stride of one. Therefore, input and output of all convolution layer have matching dimensions. So shortcuts can be used directly between each block.

our base model called ST-AttNet4 comprises four such residual block and n=45 channel size for each layer (see Fig.2 and Table 1). As is shown in Fig.2, the last residual block output is fed to temporally pooled attention module and connected softmax layer in the end. As with previous work [7,8,10], we measure the "footprint" of a model in terms of two quantities: the number of parameters and the number of multipliers in the model. our base model uses approximate 24K parameters and 2.0M multipliers. Details of our base model are shown in Table 1.

To explore the effect of model depth on performance, we design a variant of base model denoted as ST-AttNet7, which increase by 3 residual blocks without dilated convolutions. In addition to depth, we also explore the effect of model width. All models described above use n=45 channel size. We also consider a base model variant with n=65 channel size, denoted as ST-AttNet4-wide. A detailed breakdown of ST-AttNet4-narrow, our compact model, is shown in Table2. The same breakdown for our deepest model is shown in Table3.

To evaluate validity of temporally pooled attention, we construct a model, denoted as ST-Net4, which replace the temporally pooled attention module in base model with an average pooling layer as previous work.

### III. EXPERIMENTS

#### A. Datasets

We use the Google Speech Commands dataset [7] to evaluate our proposed model and baselines. For a fairer comparison, we used the same version (V1) as the baselines. The dataset consists of 64,752 recordings, containing a total of 30 words. Each recording is 1 second in duration and contains one word. 10 words are used as keywords, and the remaining 20 words are used as fillers. Following Google's implementation, our task is to discriminate among 12 classes: "yes," "no," "up," "down," "left," "right," "on," "off," "stop," "go", unknown, or silence. We use the standard list provided by the data set to divide the data set into training set, development set and test set. The specific situation is that 51088 voices are divided into the training set, 6798 voices are divided into the development set, and 6,835 voices are divided into the test set.

#### B. Experimental Setup

We use a multi-head attention mechanism at the temporally pooled attention module, with the number of heads as 5. We use tensorflow [18] to train and evaluate our model. Cross entropy is employed as the loss function. The initial learning rate is set to 0.001. After each epoch, the model will be evaluated on the development set. If the loss is not observed to decrease significantly (5%) compared to the previous epoch, then we will reduce the learning rate to 60%. In addition, when learning rate lower than $1 \times 10^{-5}$, it will be limited to $1 \times 10^{-5}$. At the same time, we force each learning rate to maintain at least 2 epochs. We use Adam algorithm [19] as

Tabel 4 The effectiveness of temporally pooled attention.

| Model | Acc. |
|---|---|
| ST-Net4 | 95.4% |
| ST-AttNet4 | 96.3% |

Table 5 Comparison of ST-AttNets and baseline models.

| Model | Acc. | Param. | Mult. |
|---|---|---|---|
| Res8-narrow[7] | 90.1% | 20K | 5.65M |
| Res15-narrow[7] | 94.0% | 43K | 160M |
| Res8[7] | 94.1% | 111K | 30M |
| Res15[7] | 95.8% | 239K | 894M |
| DS-CNN-S[8] | 94.4% | 24K | 5.4M |
| DS-CNN-M[8] | 94.9% | 140K | 19.8M |
| DS-CNN-L[8] | 95.4% | 420K | 56.9M |
| TC-ResNet8[10] | 96.1% | 66K | 1.6M |
| TC-ResNet8-1.5[10] | 96.2% | 145K | 3.3M |
| TC-ResNet14[10] | 96.2% | 137K | 3.1M |
| TC-ResNet14-1.5[10] | **96.6%** | 305K | 6.7M |
| ST-AttNet4 | 96.3% | 24K | 2.0M |
| ST-AttNet4-wide | **96.6%** | 48K | 4.1M |
| ST-AttNet7 | 96.5% | 37K | 3.3M |

the optimizer. The total number of epochs is set to 80. The mini-batch size is 100. We use the early stopping strategy [20], which means that we save the best performing model on the development set as the final model.

We have selected the model structure listed below as the baseline to verify the advantages of our proposed model in terms of parameter quantity and accuracy.

**Res8, Res8-Narrow, Res15, and Res15-Narrow** [7]. Res-variants use a residual architecture to perform keyword spotting. The number behind the Res represents the number of layers, and the narrow suffix indicates reduction of channels. Res15 is the best performance variant in Google Speech Commands Dataset

**DS-CNN-S, DS-CNN-M, and DS-CNN-L** [8]. DS-CNN uses the separable convolution as our ST-AttNet. DS-CNN-S, DS-CNN-M and DS-CNN-L denote small size, medium size and large size model, respectively.

**TC-ResNet8, TC-ResNet8-1.5, TC-ResNet14 and TC-ResNet14-1.5** [10]. They are composed of a series of one-dimensional temporal convolution residual blocks. The number after TC-ResNet represents the number of layers of the network and the 1.5 suffix indicates that the number of channels is multiplied by 1.5. TC-ResNet14-1.5 show the current state-of-the-art performance in Google Speech Commands Dataset.

Following previous work, we adopt accuracy as main metric of quality to evaluate the performance of the model. We train each model 15 for an average performance. Receiver operating characteristic (ROC) curves are employed to evaluate our model. Its x-axis is the false alarm rate, and the y-axis is the false reject rate. We scan in the threshold [0.0, 1.0], calculate the curve of a specific keyword, and then average vertically to generate the overall curve of the specific model. The smaller the area under the curve, the better the model.

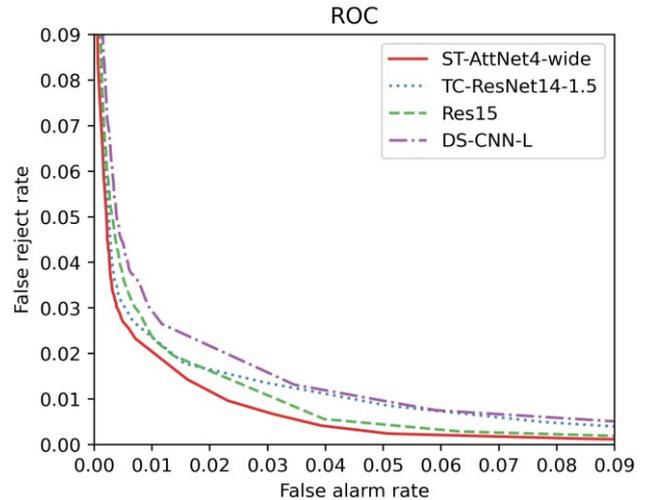

Fig.3 ROC curves for different models.

### C. Experimental Results

We investigate the effectiveness of the proposed temporally pooled attention module. The results are shown in Table 4. ST-AttNet4 is our base model. ST-Net4 adopt an AveragePool layer instead of temporally pooled attention module. We find that combining temporal convolution and temporally pooled attention is able to achieve better performance. As illustrated in Table 4, compared to ST-Net4, the accuracy of ST-AttNet4 increases by 0.9%.

In addition, we also investigate the impact of width and depth of model. The performance of our ST-ResNet variants and baselines is shown in Tabel 5. Compared to Res15 [7], the best model in ResNet variants, our ST-AttNet4 model achieves better performance with a 10× reduction of parameters and a 447× reduction of multiply operations. Compared to DS-CNN-L [8], A model that adopt a separate convolution, our ST-AttNet4 accuracy increases by 0.9% with a 17× reduction of parameters and a 17× reduction of multiply operations. Maintaining a better accuracy to TC-ResNet8-1.5 and TC-ResNet14 [10], ST-AttNet4 produces a 3× reduction in parameters and comparable multiply operations. For the best performance, ST-AttNet4-wide achieves the same accuracy of 96.6% as TC-ResNet14-1.5, while the former only requires one-seventh of parameters and fewer multiply operations.

We selected comparable models in Table 3 and plotted the ROC curve in Fig.3. The remaining models have been omitted for clarity. All curves are drawn using the best model in the testing process. The smaller the area under the curve (AUC), the better the model. The results show that the performance of ST-AttNet4-wide is better than other models.

## IV. CONCLUSIONS

In this paper, we propose a lightweight and efficient model architecture for lightweight high-performance KWS. To reduce the number of parameters and multiplications, We combine temporal convolution with depthwise separable convolution. Moreover, we introduce an temporally pooled attention module in CNN structure to capture global information. Additional comparative experiments show the effectiveness of temporally pooled attention. Futher, our ST-AttNet4-wide model achieves same accuracy of the current state-of-the-art model on Google Speech Command Dataset with only one seventh of the parameters and fewer multipliers.

ACKNOWLEDGMENT

This work is supported by National Nature Science Foundation of China (Grant No.62071039 and 61620106002 ).